\documentstyle{l-aa}

\begin{document}

   \thesaurus{
              (02.01.2;  
               02.08.1;  
               02.19.1;  
               08.06.2;  
               11.06.1)}  

   \title{Modified Artificial Viscosity in Smooth Particle Hydrodynamics}

   \author{M. Selhammar
          }

   \institute{Uppsala Astronomical Observatory\\
              Box 515\\
              S-751 20 Uppsala, Sweden\\
              Email: Magnus.Selhammar@astro.uu.se}

   \date{Received 16 September 1995; Accepted 19 November 1996}

   \maketitle

   \begin{abstract}
Artificial viscosity is needed in Smooth Particle Hydrodynamics to prevent
 interparticle penetration, to allow shocks to form and to damp post shock
 oscillations. Artificial viscosity may, however, lead to problems such as
 unwanted heating and unphysical solutions. A modification of the standard
 artificial viscosity recipe is proposed which reduces these problems. Some
 test cases discussed.
   \end{abstract}

   \section{Introduction}
Smooth particle hydrodynamics (SPH) has become an increasingly popular
 method in simulations of different astrophysical phenomena. Its Lagrangian
 formulation allows the study of large density differences. The particle
 formulation makes grids unnecessary and allows for easy implementation in
 three dimensions. A recent review of the method can be found in Monaghan
 (1992), and a technical description in for example Hernquist \& Katz
 (1989).\\
Two forms of artificial viscosity are needed in SPH, the bulk and the von
 Neumann-Richtmyer artificial viscosity respecticely. They prevent
 interparticle penetration, allow shocks to form and damp the post shock
 oscillations. They do, however, induce transfer of kinetic energy in fluid
 motions to thermal energy. Many simulations using SPH involve the
 compression of a gas, often in a gravitational collapse of an initially
 cool gas cloud. The velocities in these simulations can be highly
 supersonic, which implies that the unphysically large artificial viscosity
 may dominate the heating. This heating and deceleration of the gas prevent
 further collapse.\\
Martel et. al. (1995) have introduced a new formalism which they call
 Adaptive SPH. They use an ellipsoid kernel that adapts itself to the
 surrounding medium, and therefore avoid unnecessary use of artificial
 viscosity. The authors claim good results in cosmological collapse
 simulations.\\
In the present study a different approach is suggested. The von
 Neumann-Richtmyer artificial viscosity is restricted to supersonic
 velocities. It is therefore used only to form shocks and to prevent
 interparticle penetration for supersonic particles. Its adverse effect
 at subsonic velocities is avoided. A region under compression is described
 by a limited number of particles. The relative velocities among
 neighbouring particles in a region under compression can be supersonic
 due to the limited resolution, but despite that the gas is not expected to
 form shocks. Therefore the von Neumann-Richtmyer viscosity is restricted
 to regions that are not under compression. To avoid spurious heating in
 the subsonic regions, but prevent interparticle penetration and maintain
 the damping of post shock oscillations, the bulk artificial viscosity
 is modified. Instead of integrating the effect from the neighbouring
 particles individually, their collective effect at one point is considered.\\
The SPH method is briefly described, and the suggested changes in artificial
 viscosity are presented. These modifications of artificial viscosity have
 been tested in the simulation of a shock tube, the homologous compression
 of a gas sphere and the gravitational collapse of an initially cool gas
 sphere at rest. The results are compared with results from a standard
 formulation of artificial viscosity.

\section{The SPH Methodology}

In SPH one models a number of particles that carry the physical quantities,
 where the particles' distribution in space describes the density 
 distribution. To simulate a fluid each particle's mass is smoothed
 over a radius $r$. Following for example Hernquist \& Katz (1989),
 such a smoothed quantity at $\bf r$ be written for $N$ particles as
\begin{eqnarray}
f({\bf r}) = \oint d{\bf r}^\prime f({\bf r}) w({\bf r}^\prime-{\bf r},h) =
  \sum_j^N {\bar m_{ij}\over\bar\rho_{ij}} f({\bf r}_j) w({\bf r}_{ij},h_{ij}),
\end{eqnarray}
where $\bar m_{ij}+(m_i+m_j)/2$, $\bar\rho_{ij}=(\rho_i+\rho_i)/2$, 
$\bar h_{ij}=(h_i+h_j)/2$ and ${\bf r}_{ij}={\bf r}_i-{\bf r}_j$. The smoothing 
kernel, $w$, has the property
\begin{eqnarray}
\oint d{\bf r} w({\bf r},h)=1.
\end{eqnarray}
The kernel used here is the spline kernel from Monaghan and Lattanzio
 (1985):
\begin{equation}
w({\bf r},h) = {1\over \pi h^3} \left\{
  \begin{array}{ll}
    1 - {3\over2}\left({r\over h}\right)^2 + 
        {3\over4}\left({r\over h}\right)^3 & 0\le{r\over h}\le1 \\
    {1\over4}\left(2-{r\over h}\right)^3 & 1\le{r\over h}\le2 \\
    0 & {r\over h}\ge2 \\
  \end{array}
\right.
\end{equation}
This is a smooth kernel with compact support over the radius $2h$ around
 the particle. The smoothing length $h$ is varying in time and updated
 every iteration to keep the interactions with other particles to a
 specific number. They are called the particle's neighbours, and the
 number of neighbours for each particle in the tests in this paper is 64.\\
In the SPH formulation there are different forms of the discretization of
 the Navier Stokes equations. Here the expressions of Hernquist \& Katz
 (1989) are used, i.e.
\begin{equation}
\left\{
  \begin{array}{l}
    {d{\bf v}_i \over dt} = - \sum_j^N \bar m_{ij} 
        \bigg({\bar P_{ij}\over\bar\rho_{ij}^2} + 
        \Pi_{ij}\bigg) \nabla_i w_{ij} \\
    {d u_i \over dt} = - {1\over2} \sum_j^N \bar m_{ij} 
        \bigg({\bar P_{ij}\over\bar\rho_{ij}^2} + \Pi_{ij}\bigg) 
        {\bf v}_{ij} \cdot \nabla_i w_{ij} \\
  \end{array},
\right.
\end{equation}
where ${\bf v}_{ij}={\bf v}_j-{\bf v}_i$ and $\bar P_{ij}=(P_i+P_j)/2$. 
The continuity equation is automatically satisfied due
 to the Lagrangian formulation, and the density calculated from Eq. (1)
 becomes
\begin{equation}
\rho_i = \sum_j^N\bar m_{ij}w_{ij}.
\end{equation}
The standard artificial viscosity term, $\Pi_{ij}$, is defined
\begin{equation}
\left\{
  \begin{array}{l}
    \Pi_{ij} = {1\over\bar\rho_{ij}}(-\alpha\mu_{ij}\bar c_{ij} + 
        \beta\mu_{ij}^2) \\
    \mu_{ij}=\bar h_{ij} {{{\bf r}_{ij}\cdot{\bf v}_{ij}}\over
        {r_{ij}^2+\nu^2}} \delta_{ij} \\
    \delta_{ij} = \left\{
      \begin{array}{lll}
        1 & \mbox{if} & {\bf r}_{ij}\cdot{\bf v}_{ij}<0 \\
        0 & \mbox{if} & {\bf r}_{ij}\cdot{\bf v}_{ij}>0 \\
      \end{array}
    \right. \\
  \end{array}
\right. ,
\end{equation}
where $\bar c_{ij}=(c_i+c_j)/2$. The first and second term in the expression 
of $\Pi_{ij}$ in Eq. (6)
 represents the bulk and the von Neumann Richtmyer artificial viscosity
 respectively. The constant $\nu=0.01h$ is a fudge parameter to prevent the
 artificial viscosity to become too large. The artificial viscosity is
 used only when ${\bf r}_{ij}\cdot{\bf v}_{ij}<0$, that is when two particles 
are approaching each other.
 To close the system, the pressure is defined as $P+(\gamma-1)\rho u$, where 
$u$ is the thermal energy density and $\gamma$ the adiabatic index. The particles'
 quantities are updated using a standard leapfrog integrator with the
 time step
\begin{equation}
\delta t = {\rm Min} [\delta t_i] = \frac{\varepsilon h_i}{v_i + c_i + 1.2 (\alpha c_i + \beta \mu_{i,max})}, 
\end{equation}
where $\epsilon=0.3$  is the Courant factor to stabilize the integration, 
and $\mu_{max}$ the maximum $\mu$ from the interactions with the other particles. 
Since all particles are integrated with the same time step, the smallest time
 step from all particles is used in the integration. In the leapfrog
 integrator the velocity and internal energy density are integrated at
 half time steps, $t_{n-1/2},t_{n+1/2},...$, while the position is integrated 
at whole time steps, $t_n,t_{n+1},...$, as
\begin{equation}
  \left\{
    \begin{array}{l}
      {\bf v}_{i,n+1/2} = {\bf v}_{i,n-1/2} + {d{\bf v}_i\over{dt}}_n \delta t \\
      u_{i,n+1/2} = u_{i,n-1/2} + {du_i\over{dt}}_n \delta t \\
      {\bf r}_{i,n+1} = {\bf r}_{i,n} + {\bf v}_{i,n+1/2} \delta t \\
    \end{array}
  \right. .
\end{equation}
The viscous acceleration terms in Eq. (4) scales with $\alpha=\beta=1$ as
\begin{equation}
  \left\{
    \begin{array}{l}
      {dv\over dt}_\alpha \propto vc \\
      {dv\over dt}_\beta \propto v^2 \\
    \end{array}
  \right. .
\end{equation}
The artificial viscosity can therefore lead to undesirable effects,
 because the velocity differences are smoothed on a time scale of
 roughly $h/(c+v)$. The velocities in the model will therefore be smoothed
 out unless it expands or if there is some driving mechanism such as
 gravitation. To conserve the energy the particles are heated, which
 may be unphysical. The heated gas may reach an equilibrium state
 earlier than expected. A way of preventing interparticle penetration
 without unnecessary heating of the gas could therefore be useful, and
 this implies that there is a need to restrict the artificial viscosity
 to the shocks as much as possible.

\section{Restrictive Use of von Neumann-Richtmyer Artificial Viscosity}
In SPH the formation of shocks is mostly an effect of the von
 Neumann-Richtmeyer viscosity. To avoid the undesirable effects in the
 subsonic region, I propose to restict the use of it to those regions.
 This modification will allow shocks to form, and prevent interparticle
 penetration at supersonic velocities.\\
Consider a gas cloud in a spherical symmetric gravitational collapse.
 Suppose the physical viscosity is small, so that the compression can
 be assumed to be adiabatic. When the model has reached an equilibrium,
 the pressure force that prevents further gravitational compression
 balance the gravitational force. If the cloud is warm, the forces may
 balance each other even in its initial state. But if on the other hand
 the cloud is initially cool, it must be compressed to gain the required
 thermal energy density, perhaps by orders of magnitude. In many cases
 this is not possible with standard artificial viscosity, Eq. (6), due
 to poor resolution. If the cloud in the example above has a radius of
 unit length and is modelled with $10^3$ particles, the mean interparticle
 distances are about 0.1. In the spherical compression the particles at
 different radii therefore have supersonic relative velocities if the gas
 is cool enough. Standard artificial viscosity, Eq. (6), will decelerate
 and heat the gas, and prevent further compression.\\
Since $\rho \propto h^{-3}$ the time derivative for any particle can be written
\begin{equation}
\dot\rho=-{3\rho\dot h\over h}.
\end{equation}
This relation can be used to decide whether a particle follows the fluid
 or if artificial viscosity is necessary. Consider two particles with a
 separation of $r$ moving towards each other with a speed of $\dot r$ they
 follow the fluid in the neighbourhood, the neighbourhood is under
 compression and the particles' relative velocity will approximately satisfy
\begin{equation}
{\dot r\over r}\approx{\dot h\over h}.
\end{equation}
If two particles are approaching each other at a velocity exceeding the
 sound speed, that is if
\begin{equation}
{\bf r}_{ij}\cdot{\bf v}_{ij}<0\quad and \quad v_{ij}>\bar c_{ij},
\end{equation}
artificial viscosity is necessary to prevent interparticle penetration.
 I therefore propose to restrict the von Neumann-Richtmyer artificial
 viscosity as
\begin{equation}
  \left\{
    \begin{array}{l}
      \Pi_{vNR,ij}={\beta\mu^2_{ij}\over\bar\rho_{ij}} \\
      \mu_{ij} = \bar h_{ij}{{\bf r}_{ij}\cdot{\bf v}_{ij} \over
          r_{ij}^2 + \mu^2_{ij}} \delta_{ij} \\
      \delta_{ij} = \left\{
        \begin{array}{lll}
          1 & \mbox{if} & \left\{
            \begin{array}{l}
              {\bf r}_{ij}\cdot{\bf v}_{ij}<0 \\
              v_{ij}>c_{ij} \\
              {\dot{\bar\rho}_{ij}\over3\bar\rho_{ij}}r_{ij} > 
                  v_{ij}-\bar c{ij} \\
            \end{array}
          \right. \\
          0 & \mbox{otherwise} & \\
        \end{array}
      \right.
    \end{array}
  \right. .
\end{equation}
  
\section{Modified Bulk Viscosity}
A smoothed quantity can be calculated at any point in the fluid by
 Eq. (1). The smoothed velocity at ${\bf r}_i$ is
\begin{equation}
\Delta{\bf v}^\prime_i = \sum_j^N\Delta{\bf v}^\prime_{ij} =
    \sum_j^N{\bf v}_j{\bar m_{ij}\over\bar\rho_{ij}}w_{ij}.
\end{equation}
If this point coincides with a position for a particle, the smoothed
 and the particle's individual velocity will be different in general.
 This difference is used to construct a modified bulk viscosity. Benz
 (1990) suggests that this quantity could be used when integrating the
 position to prevent interparticle penetration. It is true that it
 prevents interparticle penetration, but it unfortunately introduces
 conservation problems when the particles are not moved at their
 individal velocity. If the contribution from particle $i$ subtracted,
 it does however say something about the fluid around the particle. The
 smoothed velocity at $r$ can then be redefined as
\begin{equation}
\Delta{\bf r}^\prime_i = {1\over1-w_i}\left(
    \sum_j^N{\bf v}_j{\bar m_{ij}\over\bar\rho_{ij}}w_{ij} -
    {\bf v}_i{\bar m_{ii}\over\bar\rho_{ii}}w_{ii}\right),
\end{equation}
where $w_{ij}=w(r_{ij}/\bar h_{ij}=0)=1/\pi h_{ij}^3$.
 The denominator scales the
 expression so that $\Delta{\bf v}_i={\bf v}_i$, if for all $j$
 ${\bf v}_j={\bf v}_i$.\\
In the standard formulation of the bulk artificial viscosity particle $i$
 interacts with each neighbour separately, where the acceleration and
 time derivative of the internal energy is added to the particle as
 described in Sect. 2. This introduces problems described in Sect. 3. 
The individual velocity of particle $i$ can then be seen as a
 deviation from the smoothed value at the particle's position and the
 artificial viscosity as a correction to the individual velocity. I
 propose a modified bulk viscosity to replace the standard bulk viscosity,
 where the fluid around the particle is considered from a collective
 contribution from the neighbours in one single interaction. This
 modified bulk artificial viscosity is defined as
\begin{equation}
  \left\{
    \begin{array}{l}
      {d{\bf v}_i\over dt} = - N_n \eta {c_i\hat{\bf v}_i\over h_i}
          \sqrt{{\bf v}_i\cdot\Delta{\bf v}_i}\delta_i \\
      \delta_i = \left\{
        \begin{array}{lll}
          1 & \mbox{if} & {\bf v}\cdot\Delta{\bf v}_i < 0 \\
          0 & \mbox{if} & {\bf v}\cdot\Delta{\bf v}_i > 0 \\
        \end{array}
      \right. \\
    \end{array}
  \right. ,
\end{equation}
where $N_n$ the number of neighbours. The constant $\eta$ around unity, and
 used in the same way as the constants $\alpha$ and $\beta$ in Eq. (6). This
 interaction can be seen as if the particle interacts with a virtual
 particle with the smoothed velocity $\Delta{\bf v}_i$, have a mass of
 $N_n m_i$ and lies at a distance of $h$ the direction of ${\bf v}_i$.
 This expression thus becomes similar to Eq.(6) and
 affects the same velocity regime. The difference is that in Eq. (16)
 the collective contribution from all neighbours is considered in one
 single interaction.\\
Now consider the integration of the velocitites and internal energy
 density from time $t_{n-1/2}$ to $t_{n+1/2}$ with the time step $\delta t$.
From Eq. (8) the velocity for particle $i$ at $t_{n+1/2}$ is
\begin{equation}
{\bf v}_{i,n+1/2} = {\bf v}_{i,n-1/2} + {d{\bf v}_{i,n}\over dt} \delta t,
\end{equation}
which gives the change in kinetic energy for particle $i$:
\begin{equation}
\matrix{
  \Delta T_i = {m_i\over2}\left({\bf v}_{i,n+1/2}\cdot{\bf v}_{i,n+1/2} -
      {\bf v}_{i,n-1/2}\cdot{\bf v}_{i,n-1/2}\right) =\cr
  \quad\quad m_i{\bf v}_{i,n-1/2}\cdot{d{\bf v}_{i,n}\over dt} +
      {m_i\over2}{d{\bf v}_{i,n}\over dt}\cdot{d{\bf v}_{i,n}\over dt}
      \delta t^2}.
\end{equation}
If $\Delta U_i = m_i\left(du_i/dt\right)\delta t$ is the change in
 internal energy for the particle it is possible
 to conserve the energy, that is require that $\Delta T_i + \Delta U_i = 0$.
 The time derivative
 of the internal energy for the particle is then consequently defined as
\begin{equation}
{du_{i,n}\over dt} = {\Delta U_i\over{m_i\delta t}} = 
    - {\bf v}_{i,n-1/2}\cdot{d{\bf v}_{i,n}\over dt}\delta t -
    {1\over2}{d{\bf v}_{i,n}\over dt}\cdot{d{\bf v}_{i,n}\over dt}\delta t^2
\end{equation}
to conserve the total energy.\\
The artificial also must prevent particle penetration. The modified
 artificial viscosity is calculated with respect to an integrated mean
 of the neighbours. Therefore a particle will move less than $\epsilon h$
 from the definition of the time step, Eq. (7), regardless of the neighbours'
 individual sound speed and individual velocities. This should be
 compared with the distance to the closest which are approximately
 one $h$ with 64 neighbours. Since the neighbours have no identity, this
 may lead to penetration with a few particles which have a sufficient
 deviation from the integrated mean.\\
If the particles $i$ and $j$ are each others neighbours and that their
 other neighbours give the same contribution to their respective
 velocities, one concludes from Eq. (15) and (16) that the impulse is
 conserved. Their other neighbours do, however, not give the same
 contribution, due to the limited resolution. Tests of self gravitating
 rotating disks show that the angular momentum and impulse are well
 conserved. \\

\section{Tests}
Any form of artificial viscosity must be able to form and propagate a
 shock. The modification of the artificial viscosity, Eq. (13) and
 (16), is tested in a shock forming test and compared with the standard
 artificial viscosity, Eq. (6),  introduced by Monaghan and Gingold 
(1983). The ability of the modified artificial viscosity to compress
 the gas without viscous deceleration and heating has also been tested
 in a homologous compression of a gas sphere. The test constructed by
 Evrard (1988) is used to study the differences between the artificial
 viscosities in a gravitational collapse of an initially cool gas cloud.\\
The number of particles is varied between 8192 and 16384, and the number
 of neighbours for each particle is 64, so that h is varying in time and
space. In
 the equation of state the adiabatic index is $\gamma=5/3$ to model an ideal
 monoatomic gas. Dimensionless units are used to keep the quantities
 in the model around unity, where the gravitational constant $G=1$. In a
 model the total mass $M=1$, the typical length $L=1$ and time $T=1$, which
 relates to the gravitational constant as $G=L^3/MT^2$. The real quantities of
 the model can be calculated by inserting the corresponding quantities
 in the desired unit system.\\

\begin{figure*}
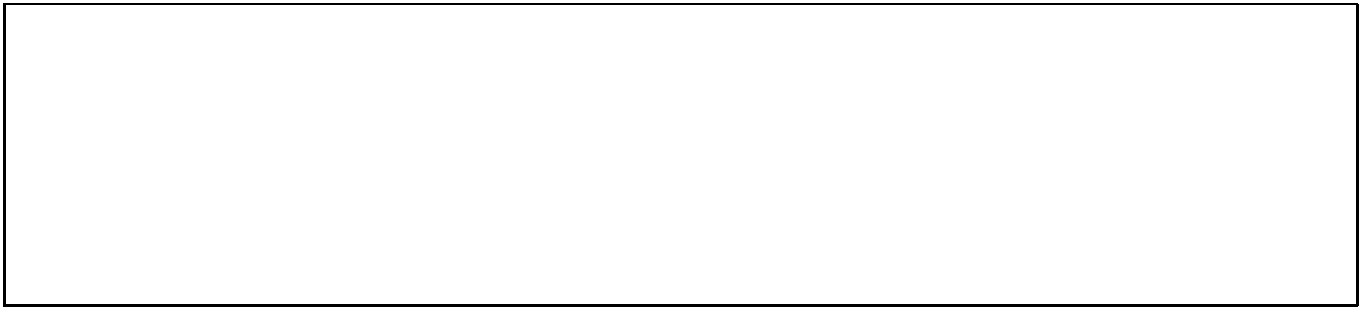

\picplace{4.0cm}
\caption{
A shock in a shock tube with standard artificial viscosity, Eq.
 (6), which was formed by the density discontinuity described in Eq.
 (20). Fig. 1a shows at time $t=20$ the velocity distribution, 1b the
 pressure and 1c the density distribution.}
\end{figure*}
\begin{figure*}
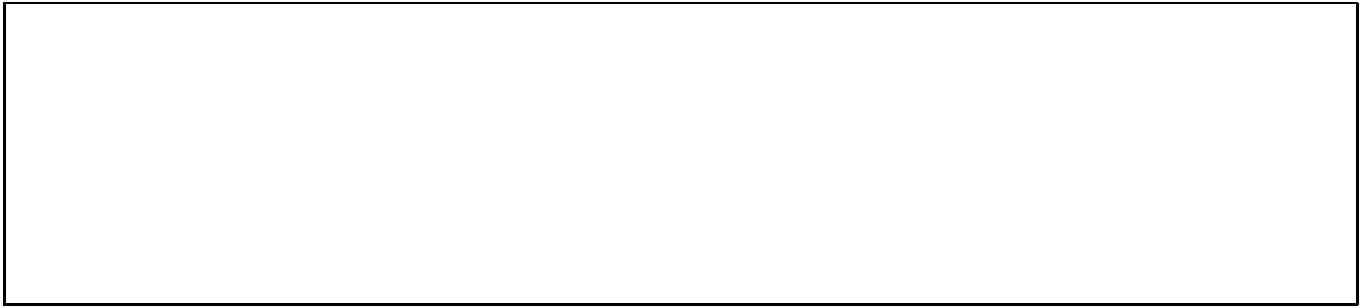

\picplace{4.0cm}
\caption{
Same as Fig. 1, but with modified artificial viscosity, Eq. (13)
 and (16).}
\end{figure*}

\subsection{Shock Formation Test}
A box with dimensions $d_x\times d_y\times d_z=1\times1\times9$ is used.
 Periodic boundary conditions are
 applied in the $x$- and $y$-directions. At the ends of the tube, $z=-5$ and
 $z=4$, no boundary conditions were applied, so the particles are allowed
 to move away from the tube. Their velocities are however low compared
 with the velocity of the shock and do not affect the shock model. The
 16384 particles are ordered in a cubic centered grid, such that 8192
 particles are distributed at $-5\le z<-4$, and the rest at $-4\le z\le4$. With
 a total mass of 1.0 mass units the density distribution is
\begin{equation}
  \rho=\left\{
    \begin{array}{lll}
      1/4 & \mbox{where} & -5\le z<-4 \\
      1/16 & \mbox{where} & -4\le z\le4 \\
    \end{array}
  \right.
\end{equation}
The initial thermal energy density is set to 0.01, and the particles
 have no initial velocity. A shock is formed at the discontinuity at
 $z=-4$, which propagates to the right. This is a rather weak shock, which
 is a better test than a strong shock. The reason is that here the
 particle velocities are not much larger than the sound speed, because
 if they are the standard and modifies artificial viscosity become
 similar. Fig. 1 shows the shocks at $t=20$ with the standard artificial
 viscosity, Eq. (6), and Fig. 2 using modified artificial viscosity,
 Eq. (13) and (16). A comparison between Fig. 1 and 2 shows that the
 shocks formed by the two versions of artificial viscosity are similar,
 so that the modified artificial viscosity is able to work in the same
 way as the standard artificial viscosity.\\
\begin{figure}
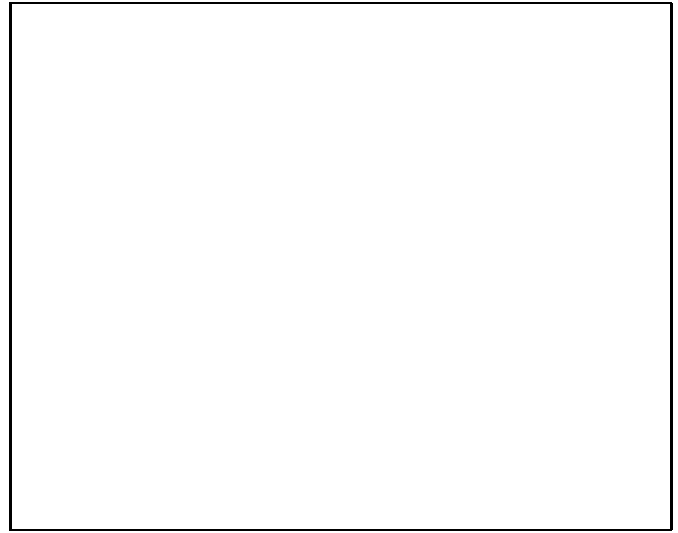

\picplace{7.0cm}
\caption {
The energy curves for the homologous compression of a gas with
 different forms of artificial viscosity. The solid line represents
 the compression without any artificial viscosity at all, the dashed
 the modified artificial viscosity, Eq. (13) and (16), and the dotted
 the standard artificial viscosity, Eq. (6). The curves that decline
 in the beginning represent the kinetic energies for the three models,
 those that rise represent the internal and the uppermost straight
 curves represent the total energy.}
\end{figure}
\begin{figure*}
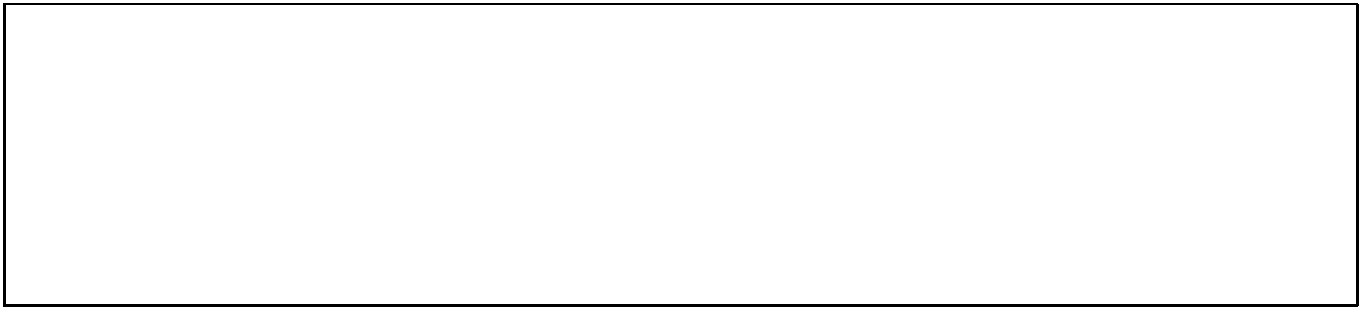

\picplace{4.0cm}
\caption {
The density distribution for the homologous compression of the
 sphere at maximum compression without any artificial viscosity in 4a,
 with standard artificial viscosity in b, Eq. (6), and modified
 artificial viscosity, Eq. (13) and (16) in 4c. The results in Fig.
 4a and 4c are plotted with the same scale, while another scale must
 be used in 4b. The solid line in Fig. 4a and c represent the zeroth
 order theoretical estimate from Eq. (26) at maximum compression where
 the radius is 0.029 and the density 10000.}
\end{figure*}

\subsection{Homologous Compression of a Gas Sphere}
The ability of the modified artificial viscosity, Eq. (13) and (16), to
 handle compression of a gas realistically is tested. Initially 8192
 particles are distributed unifomly in a sphere with radius $R_{init}=1$ on a
 slightly disturbed cubic centered grid. There are no boundary conditions
 applied, but there is an initial velocity distribution directed towards
 the origin according to
\begin{equation}
{\bf v}_{init}\left({\bf r}\right)=-V_0{{\bf r}\over R_{init}},
\end{equation}
where $V_0=2$ is a constant and $\bf r$ the position in the sphere.
 The particles are initially isothermal with a thermal energy density
 of $u_{init}=0.001$, and with a total mass of $M=1$, which gives an
 initial density of $\rho_{init}=3/4 \pi$. The sum of
 the total kinetic, $E_{kin}$, and thermal, $E_{th}$, energies is
\begin{equation}
  \begin{array}{l}
    E_{tot} = E_{kin} + E_{th} = \int_0^M{v^2\over2}dm + uM= \\
    \quad\quad\int_0^R{4r^2\over2}4\pi r^2\rho dr + \mu M = 
        {6\over5}+0.001=1.201 \\
  \end{array}
\end{equation}
Assume that the compression is adiabatic, so that Poisson's equation,
\begin{equation}
P=K\rho^\gamma,
\end{equation}
is valid. From the equation of state, $P=(\gamma-1)u\rho$, and the
 adiabatic index, $\gamma=5/3$, it follows that
\begin{equation}
K = (\gamma-1)\mu_{init}\rho_{init}^{\gamma-1}=1.73\cdot10^{-3}
\end{equation}
If it is assumed that all kinetic energy is converted to thermal energy
 at maximum compression, the thermal energy density at this point is
\begin{equation}
u^\prime = {E_{tot}\over M}=1.201.
\end{equation}
This gives a density and radius of the compressed gas sphere as
\begin{equation}
\matrix{
  \rho^\prime = \left[{K\over(\gamma-1)u^\prime}\right]^{1\over1-\gamma}\cr
  R^\prime = \left({3M\over4\pi\rho^\prime}\right)^{1\over3}=0.029\cr}
\end{equation}
This zeroth order approximation is useful to compare with calculations
 with different forms of artificial viscosity. The initial conditions
 also have the advantage that no artificial viscosity is needed to
 prevent interparticle penetration. A small initial pressure is
 sufficient to decellerate the particles to zero and prevent them
 to move through the origin. The results from the test with modified
 artificial viscosity can therefore not only be compared with the
 analytical approximation Eq. (25), but also with a model without
 any artificial viscosity. The energy curves from such a comparison
 are shown in Fig. 3. In Fig. 4 the density distributions at time $t=0.5$
 are compared.\\
The heating in the case with standard artificial viscosity, Eq. (6),
 starts immediately, because it depends on the relative velocities,
 while the heating without artificial viscosity and with the modified
 artificial viscosity is negligable until $t\approx0.3$. The modified
 artificial viscosity has a little less steep energy curve compared with
 the case without artificial viscosity, but gives an almost a compressed
 gas as without artificial viscosity as seen in Fig. 4. The true density
 distribution is not known, but the zeroth order approximation from
 Eq. (25) gives approximately the same size as the model without
 artificial viscosity.\\

\subsection{The Evrard Gravitational Collapse}
\begin{figure}
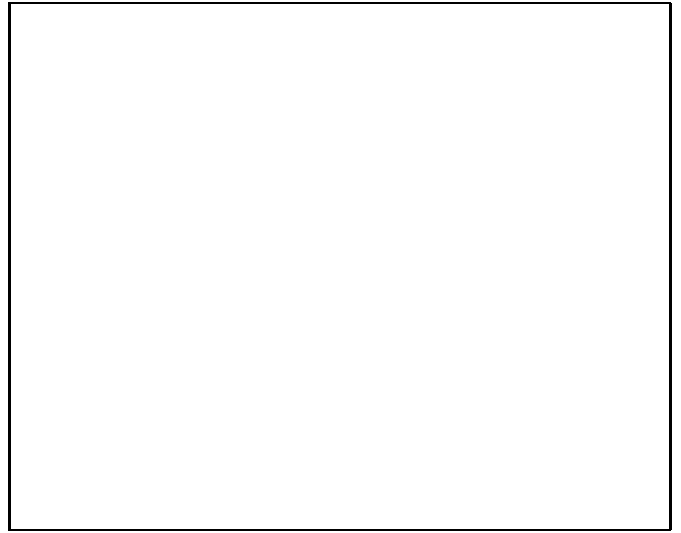

\picplace{7.0cm}
\caption {
The energy curves for the Evrard gravitational collapse using
 standard artificial viscosity, Eq. (6), represented by a dashed
line, and modified artificial viscosity, Eq. (13) and (16),
represented by a solid line. The uppermost curves represent the
internal thermal energy, the next the kinetic, the straight line
the total and the two curves below the others' the potential energy.
 The solid line is from an accurate one-dimensional PPM calculation
at $t=0.77$ from Steinmetz \& M\"uller (1993).}
\end{figure}
\begin{figure*}
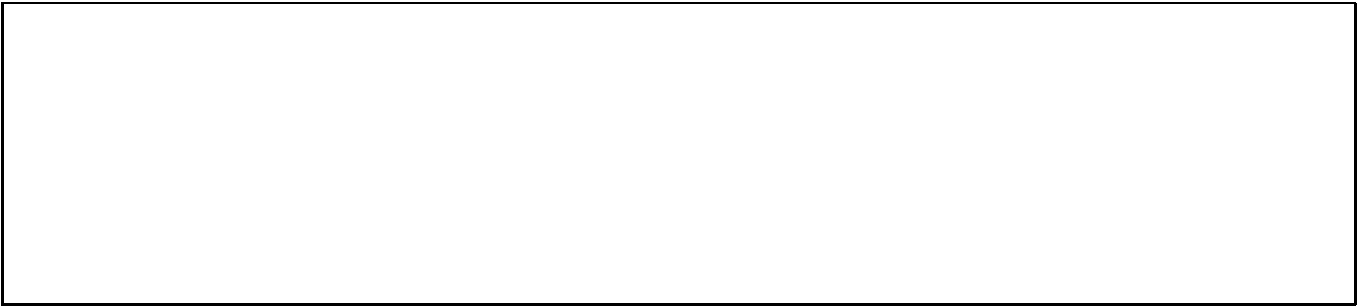

\picplace{4.0cm}
\caption {
The velocity, density and pressure distributions for the Evrard
 gravitational collapse using standard artificial viscosity, Eq. (6).
 The radial velocity, density and pressure are plotted at t=0.8. The
 crosses are values at t=0.77 are results from an accurate
 one-dimensional PPM calculation from Steinmetz \& M\"uller (1992).}
\end{figure*}
\begin{figure*}
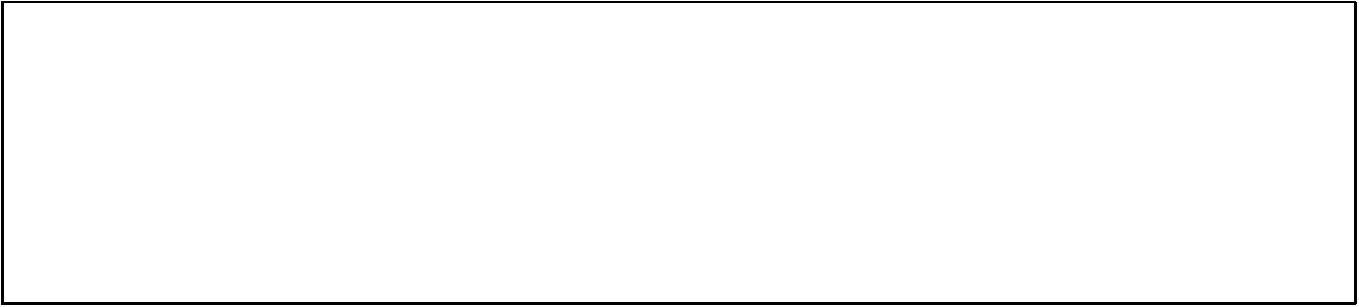

\picplace{4.0cm}
\caption {
Same as Fig. 6, but with modified artificial viscosity,
 Eq. (13) and (16).}
\end{figure*}

The initial conditions are those from Evrard (1988), which is a
 isotherm sphere at rest with a thermal energy density of $u=0.05$,
 radius $R=1$ and mass $M=1$. Initially 8192 particles are distributed
 uniformly in a sphere with radius $R=1$ on a slightly disturbed cubic
 centered grid with a density distribution of 
\begin{equation}
\rho(r) = {M\over2\pi R}{1\over r}.
\end{equation}
Standard, Eq. (6), and modified artificial viscosity, Eq. (13) and
 (16) are tested with this model and compared with each other.\\
Due to the cold initial state the sphere begins to collapse. Since
 the central part is more dense than the outer payers, the collapse
 is more rapid around the origin. A high central pressure and density
 is build up, and the central parts starts to expand at $t\approx0.8$.
 Where the expanding parts meet the infalling gas, an outward propagating
 shock front forms. Eventually the gas reach a virial equilibrium.\\
The total energies are shown in Fig. 5a and b. The curves are more
 shallow in Fig. 5a, where standard viscosity was used, than in Fig.
 5b. In Fig. 6 the velocity, density and pressure distributions are
 plotted with standard artificial viscosity at $t=0.8$ when the shock is
 formed. This can be compared with the results from modified artificial
 viscosity for the same quantities which are shown in Fig. 7. The main
 difference between these models are the sharper gradients with modified
 artificial viscosity. This is an effect of the ability to compress the
 gas with modified artifial viscosity. The infalling gas is allowed to
 move inwards without decelleration until it meets the shock front.\\

\section{Conclusions}
In SPH two forms of artificial viscosity is necessary, the bulk and von
 Neumann-Richtmyer viscosity. The standard recipy does however induce
 smoothening of the velocity differences and heating of the gas. Another
 problem is that standard artificial viscosity prevents compression of
 a gas. \\
To partly overcome some of these problems, I propose a modification of
 the artificial viscosity. The von Neumann-Richtmyer artificial
 viscosity is used only in supersonic regions where the gas is not
 under compression. The bulk viscosity is modified to be calculated
 from an integrated mean of the velocities in the neighbourhood of
 the particle, instead of calculating the viscous interaction from
 each neighbour separately. Tests cases have been performed to test
 these abilitites of the proposed modified form of artificial
 viscosity, and to compare them with standard artificial viscosity. 

\begin{acknowledgements}
I wish to thank my supervisors Bengt Gustafsson and Lars Stenholm. Part
 of this work was supported by Swedish Defence Research Establishment.
 All calculations has been performed at the Center for Parallel
 Computer at KTH.
\end{acknowledgements}

\end{document}